\documentclass{article}

\input{vsolj02.sty}

\usepackage{graphicx}
\usepackage[comma,colon]{natbib}

\usepackage[hidelinks]{hyperref}

\RequirePackage[T1]{fontenc}

\def\cite{\citealt}
\setcitestyle{aysep={}}

\begin{document}
\title{Orbital period of the IW And-type star ST Cha}

\author{Taichi Kato$^1$,Naoto Kojiguchi$^1$}
\author{$^1$ Department of Astronomy, Kyoto University,
       Sakyo-ku, Kyoto 606-8502, Japan}
\email{tkato@kusastro.kyoto-u.ac.jp}

\begin{abstract}
We analyzed TESS data of the IW And-type dwarf nova
ST Cha during ordinary dwarf nova states.
We have identified an orbital period of 0.285360(1)~d.
The object was reported to be eclipsing in the past using
the data obtained in the 1970s, which is not in agreement with
the present data.  Despite the constant mean brightness,
the strength of the orbital signal varied significantly,
suggesting that the strength of the orbital signal does
not always reflect the mass-transfer rate.
During an outburst with a shoulder, we did not find
evidence of humps recurring with a period longer than
the orbital period which were recorded in V363 Lyr.
This finding strengthens the idea that V363 Lyr
is an unusual object.  We found that the strength
of the orbital signal increased after an outburst
with a shoulder.  This outburst may have changed
the state of the disk and the hot spot became
more apparent.  Such a change in the disk may have triggered
a transition from an ordinary dwarf nova-type state to
an IW And-type state and this possibility would require
further examination.
\end{abstract}

\section{Introduction}

   IW And stars are a subgroup of Z Cam type dwarf novae
[or general information of dwarf novae,
see e.g. \citet{war95book}].  Z Cam stars are
characterized by the presence of standstills in addition
to ordinary dwarf nova-type outbursts.
Standstills in Z Cam stars usually occur after
outbursts and end with fading (see e.g. \cite{szk84AAVSO}).
In IW And stars, outbursts and standstills sometimes
occur regularly (IW And-type state) and the light curves are
characterized by brightening from a standstill
(or standstill-like phase) sometimes followed by a deep dip
\citep{sim11zcamcamp1,kat19iwandtype}.
Accretion disks of the dwarf novae of this type are considered
to be near the thermal stability (see e.g.
\cite{kim20iwandmodel}).

   After recognition of the initial two prototypical
objects, IW And and V513 Cas \citep{sim11zcamcamp1},
ST Cha was the third object of this group
\citep{sim14stchabpcra,kat19iwandtype}.
\citet{kat21stcha} studied ground-based time-resolved
photometry to characterize the outburst pattern
of this object and found that the brightness level
of the shoulders (or embedded precursors) during
long outbursts is the same as the ones when the object
started brightening at the end of standstills.

   In this paper, we identified the orbital period
of this object using the Transiting Exoplanet Survey
Satellite (TESS) observations\footnote{
  $<$https://tess.mit.edu/observations/$>$.
}.  The full light-curve
is available at the Mikulski Archive for Space Telescope
(MAST\footnote{
  $<$http://archive.stsci.edu/$>$.
}).

\section{Observations}

   The TESS spacecraft observed ST Cha at a two-minutes
cadence during two segments: 
between 2019 April 24 and 2019 June 18 (segment 1) and
between 2021 April 29 and 2021 June 24 (segment 2).
We used simple-aperture-photometry (SAP) data.
The zero point of the TESS data is rather arbitrary
and these magnitudes should be regarded as relative
ones rather than zero-point calibrated ones. 
These segments contained ordinary dwarf nova states
and not IW And-type states
[see \citet{kat21stcha} for a light curve containing
both states].

\begin{figure*}
  \begin{center}
    \includegraphics[width=16cm]{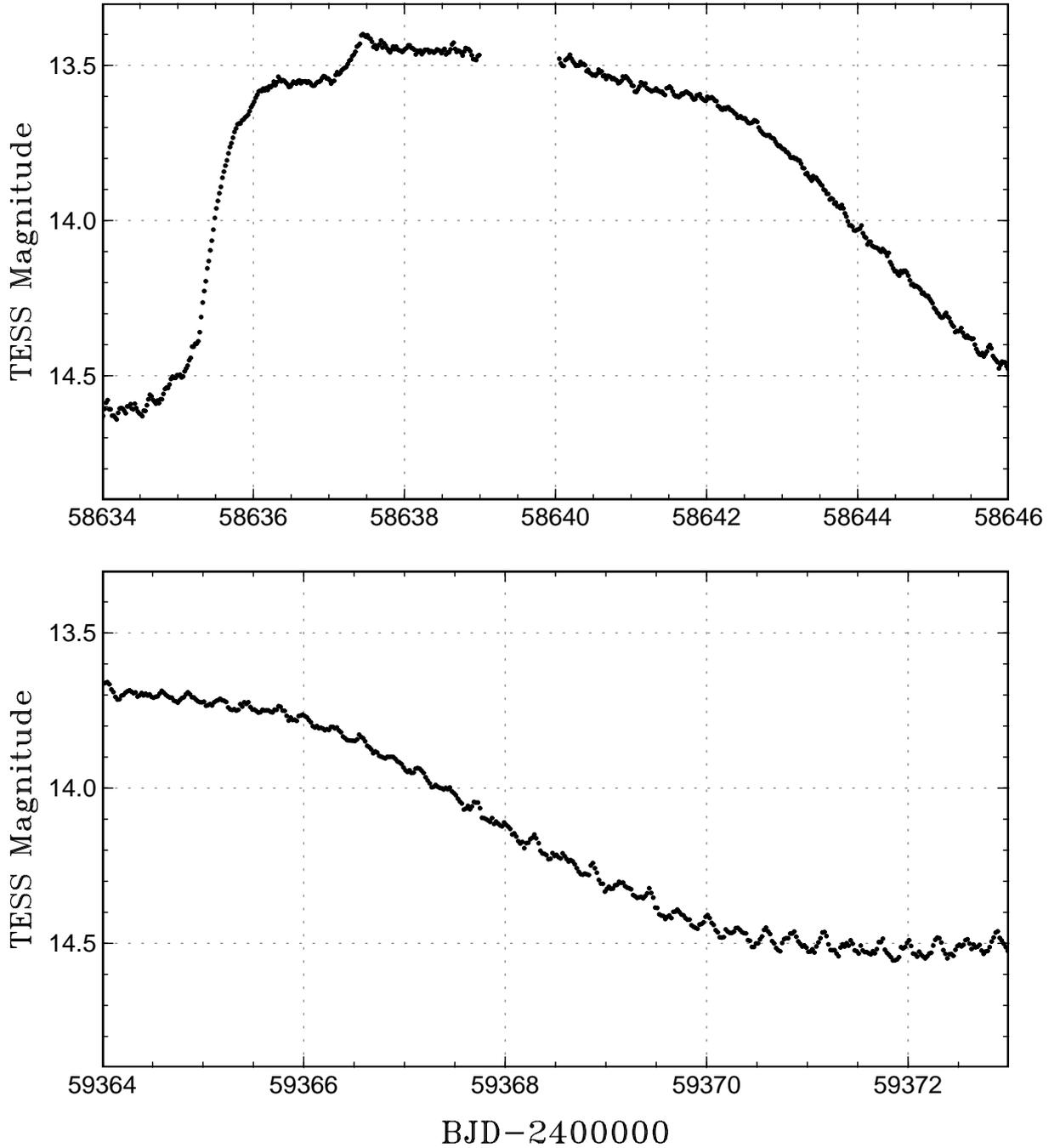}
  \end{center}
  \caption{Examples of TESS light curves of ST Cha.
  The data were binned to 0.02~d for better visibility
  of variations.
  (Upper): Long outburst with a shoulder (or embedded precursor).
  (Lower): A short segment in which orbital variations were
  directly visible.
  }
  \label{fig:stchalc}
\end{figure*}

\begin{figure*}
  \begin{center}
    \includegraphics[width=16cm]{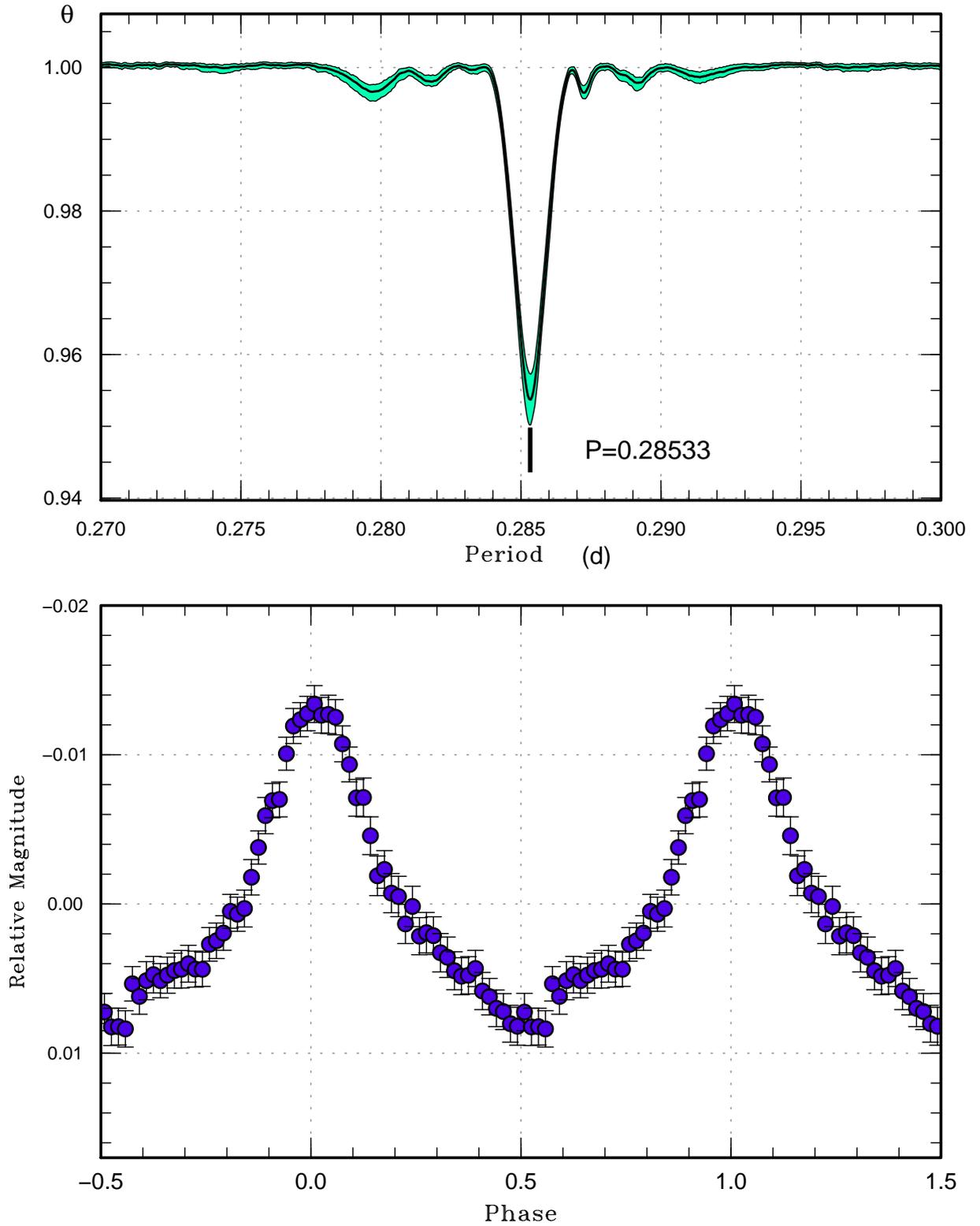}
  \end{center}
  \caption{Period analysis of the segment 1.
  (Upper): We analyzed 100 samples which randomly contain 50\% of
  observations, and performed the PDM analysis for these samples.
  The bootstrap result is shown as a form of 90\% confidence intervals
  in the resultant PDM $\theta$ statistics.
  (Lower): Orbital variation.
  }
  \label{fig:stchapdm1}
\end{figure*}

\begin{figure*}
  \begin{center}
    \includegraphics[width=16cm]{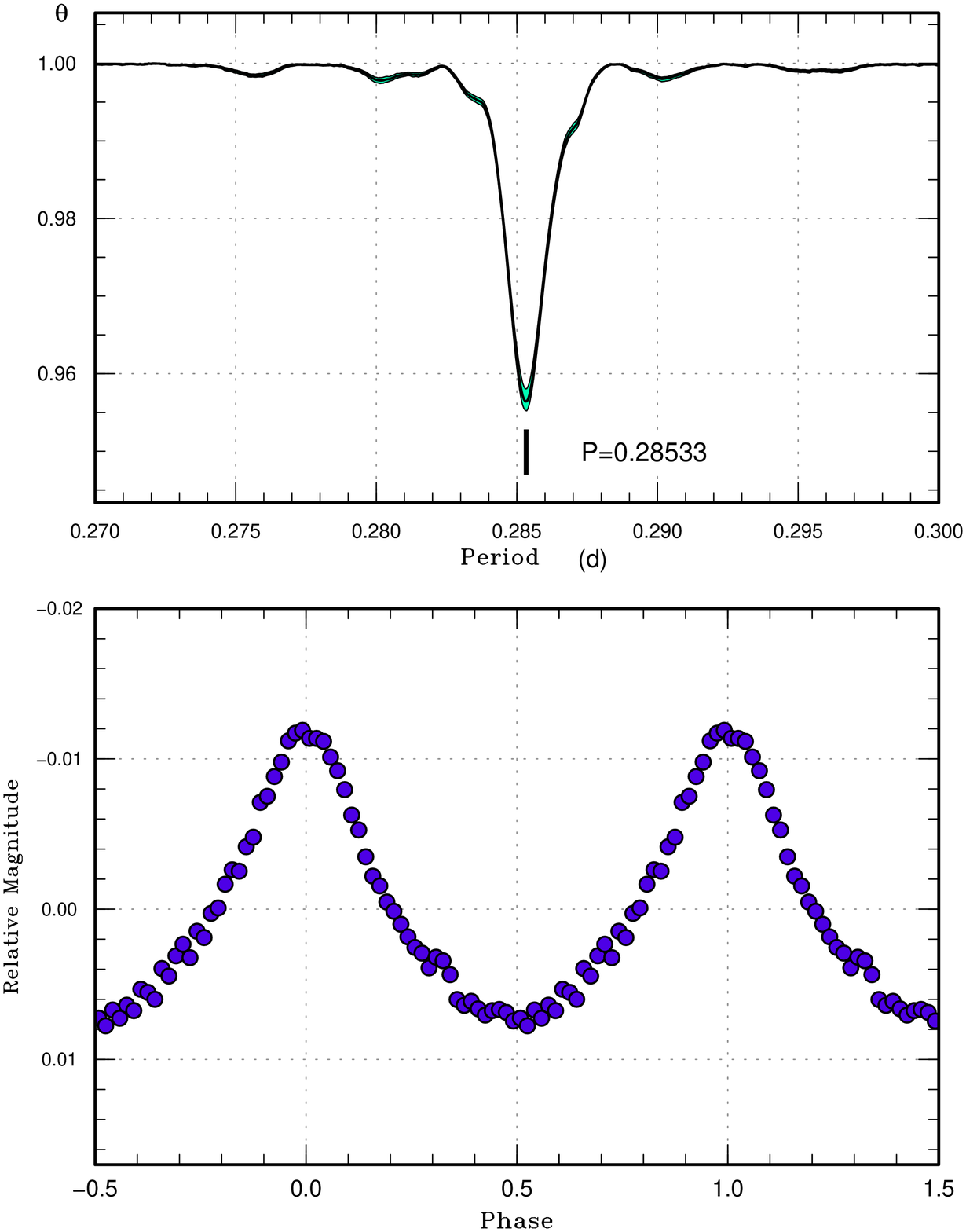}
  \end{center}
  \caption{Period analysis of the segment 2.
  (Upper): PDM analysis.
  (Lower): Orbital variation.
  }
  \label{fig:stchapdm2}
\end{figure*}

   Examples of TESS light curves are shown in figure
\ref{fig:stchalc}.  The upper panel shows a long outburst
with a shoulder in the segment 1.
The lower panel shows a short segment (from the segment 2)
in which orbital variations were directly visible.
See also upper panels of
figures \ref{fig:stchapdm1} and \ref{fig:stchapdm2}
for the entire TESS observations.

\section{Results and Analysis}

\subsection{Orbital period}

   We made period analysis using Phase Dispersion Minimization
(PDM, \cite{PDM}) after removing the trends of outbursts
by locally-weighted polynomial regression (LOWESS: \cite{LOWESS}).
The errors of periods by the PDM method were
estimated by the methods of \citet{fer89error} and \citet{Pdot2}.
Both segments 1 and 2 yielded almost the same periods
[0.28533(3)~d for the segment 1 and 0.28533(1)~d for
the segment 2].  The signals were very stable and we
identified this period as the orbital one.

\begin{figure*}
  \begin{center}
    \includegraphics[width=16cm]{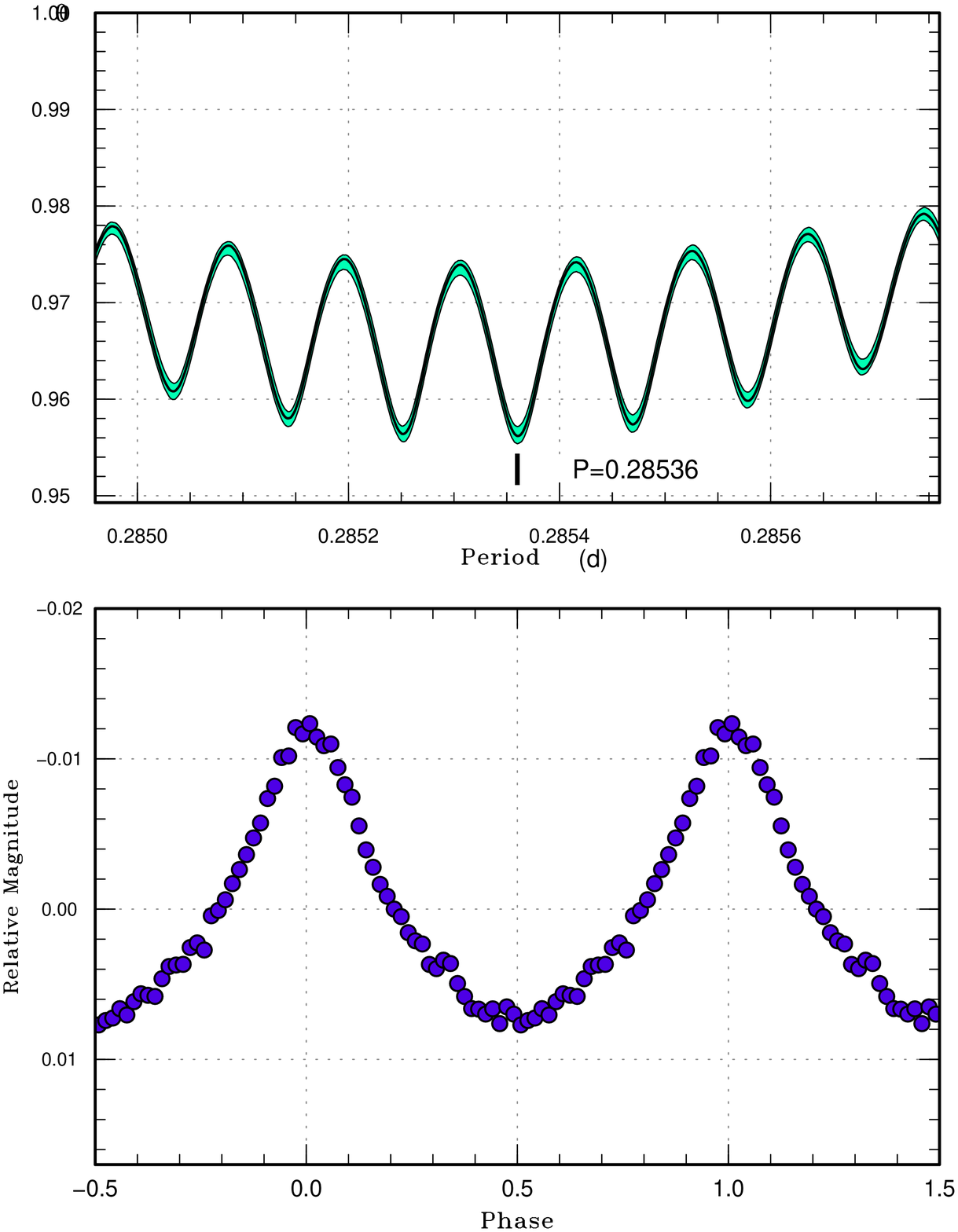}
  \end{center}
  \caption{Period analysis of all the data.
  (Upper): PDM analysis.
  (Lower): Orbital variation.
  }
  \label{fig:stchaall}
\end{figure*}

   The result of PDM analysis after combination of
the two segments is shown in figure \ref{fig:stchaall}.
A period of 0.285360(1)~d, 1$\sigma$ and 3$\sigma$
different from the periods obtained from the segments
1 and 2, respectively, has been identified as
the most likely orbital period by this study.

\subsection{Eclipses in the past?}

   The orbital period of ST Cha was suggested to be
0.285~d by \citet{ste88stcha}, who analyzed photometric
data by \citet{mau75stcha} and concluded that ST Cha
is an eclipsing binary with a period of 6.85~hours
(=0.285~d) allowing possibilities of aliases of
$P=1.9975/(2n+1)$ hours, where $n=0, 1, 2, ...$.
Combined with a low-resolution spectrum, \citet{ste88stcha}
concluded that ST Cha is either an optically selected
low-mass X-ray binary (LMXB) or a novalike cataclysmic
variable.  The TESS observations, however, did not show
any sign of eclipses and the orbital variations were
in low amplitudes ($\sim$0.02~mag).
Such low-amplitude variations would have been difficult
to detect by photographic observations by \citet{mau75stcha}
and the identity of the periods obtained by this and our
studies is likely a chance coincidence, unless ST Cha was 
an eclipsing binary in the past which stopped to be
eclipsing (such a case may not be unthinkable; there
are 10 systems among classical eclipsing binaries,
see e.g. \cite{zas12hshya}).

\subsection{Strength of orbital signal and mass-transfer rate}
   
\begin{figure*}
  \begin{center}
    \includegraphics[width=16cm]{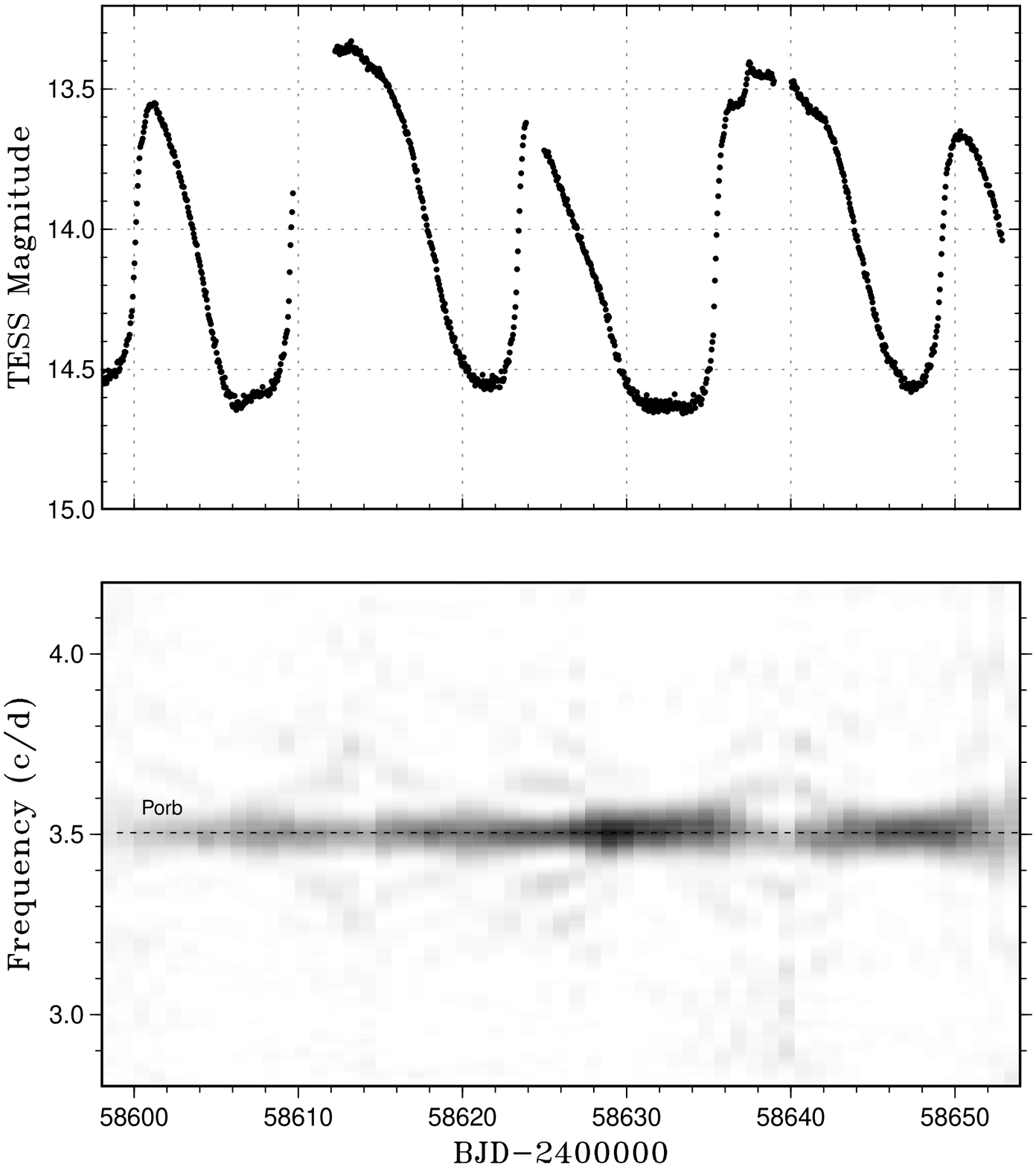}
  \end{center}
  \caption{(Upper): TESS light curve of the segment 1.
  The data were binned to 0.05~d.
  (Lower): Two-dimensional PDM analysis.
  The width of the sliding window and the time step used
  are 10~d and 1~d, respectively.  Dark colors represent signals
  (lower $\theta$ in the PDM statistics).
  The orbital signal was persistently recorded.
  }
  \label{fig:stchaspec2d1}
\end{figure*}

\begin{figure*}
  \begin{center}
    \includegraphics[width=16cm]{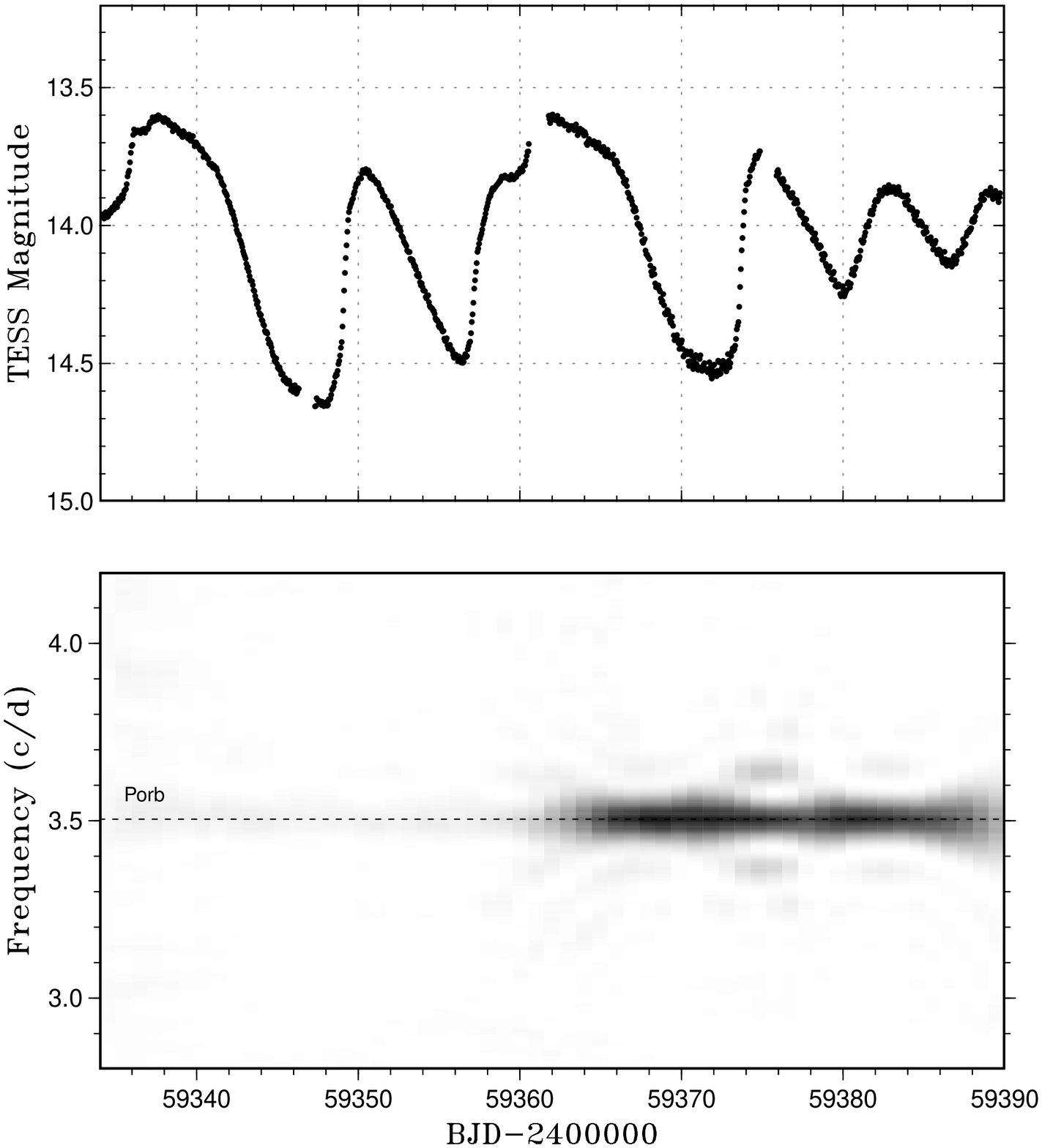}
  \end{center}
  \caption{(Upper): TESS light curve of the segment 2.
  The data were binned to 0.05~d.
  (Lower): Two-dimensional PDM analysis.
  The strength of the orbital signal varied significantly.
  }
  \label{fig:stchaspec2d2}
\end{figure*}

   We performed two-dimensional PDM analysis of the two
segments (figures \ref{fig:stchaspec2d1},
\ref{fig:stchaspec2d2}) (see \cite{kat21v363lyr}).
The orbital signal was persistently recorded during
the segment 1, but was weak during the initial half
of the segment 2.  Although the strength of orbital humps
is often considered to be an indicator the mass-transfer rate,
this result suggests that it is not always the case
since the mean brightness did not significantly varied
during the segment 2.  It may have been that the hot spot
was sometimes not formed such as by stream overflow
as suggested for SW Sex stars \citep{hel00swsexreview}.
There was no evidence of positive or negative superhumps.

\subsection{Enhanced orbital signal after outburst with shoulder?}

   \citet{kat21v363lyr} reported that humps recurring
with a period slightly longer than the orbital one
were recording by Kepler observations of the dwarf nova V363 Lyr
during a long, bright outburst associated with a shoulder.
The nature of these humps in V363 Lyr was inconclusive:
(1) V363 Lyr may have an anomalously undermassive secondary
and this object is an SU UMa star above the period gap or
(2) The humps were a result of deformation of the disk
at the tidal truncation radius.  In the case of ST Cha,
an outburst associated with a shoulder in the segment 1
did not show any sign of the humps similar to V363 Lyr.
The case of V363 Lyr looks like to be special in that
it showed brightness increase by 0.5~mag after the shoulder
in contrast to ST Cha.  After an outburst associated
with a shoulder in the segment 2 (figure \ref{fig:stchaspec2d2}),
however, the orbital signal became stronger.  It may be
that the outburst with a shoulder changed the state of the disk
and the hot spot may have become more apparent.
Such a change in the disk may have triggered
a transition from an ordinary dwarf nova-type state to
an IW And-type state and this possibility would require
further examination.

\subsection{Are IW And-type states brighter than dwarf
   nova-type states?}

\begin{figure*}
  \begin{center}
    \includegraphics[width=16cm]{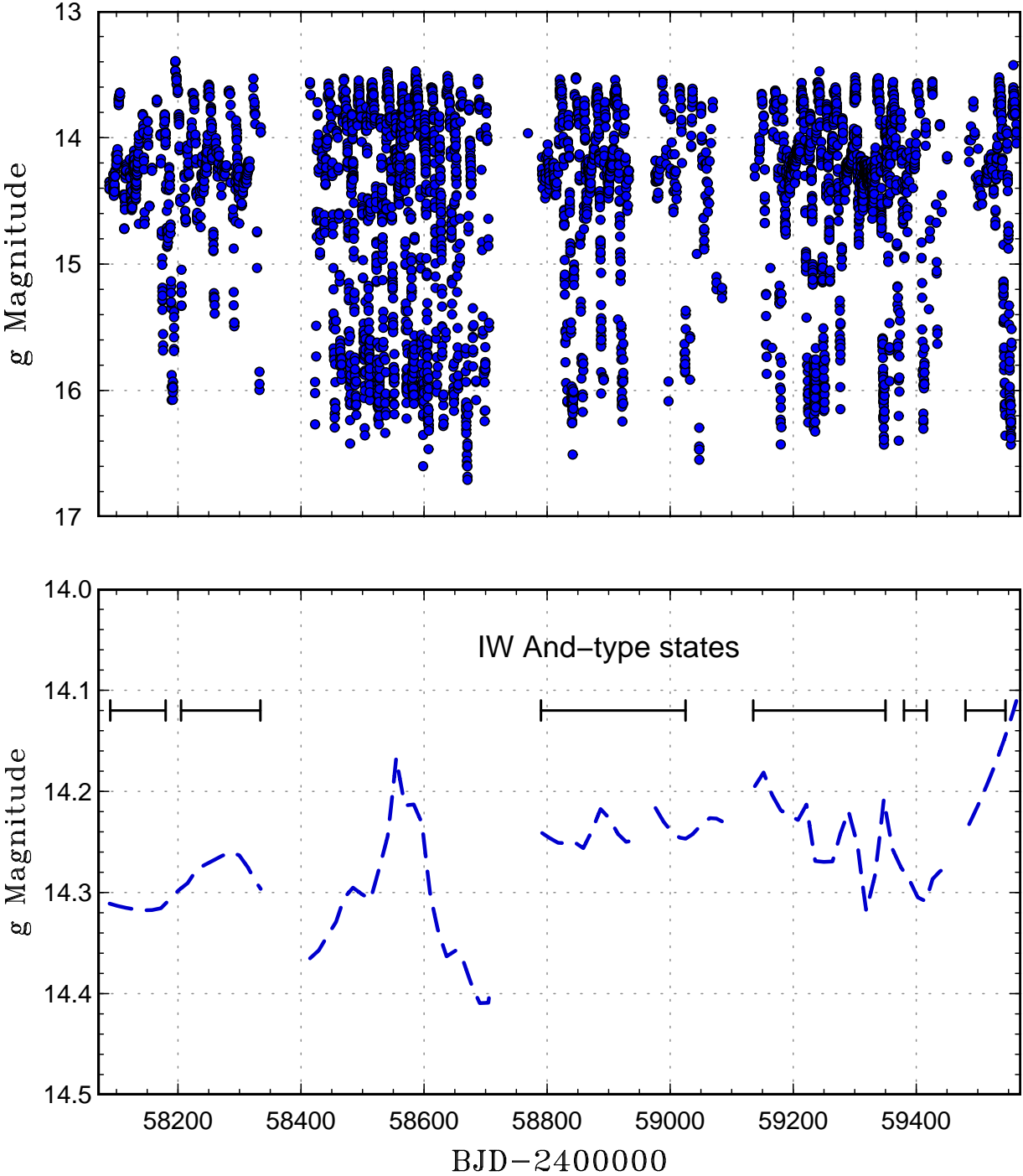}
  \end{center}
  \caption{(Upper): ASAS-SN $g$-band light curve of ST Cha.
  (Lower): Trends determined by LOWESS.
  Horizontal marks represent IW And-type states.
  }
  \label{fig:stchaave}
\end{figure*}

   We used All-Sky Automated Survey for Supernovae (ASAS-SN:
\cite{ASASSN,koc17ASASSNLC}) $g$-band observations to see
whether IW And-type states are brighter than dwarf
nova-type states.  The method is the same as described
in \citet{kat21ixvel} using LOWESS to obtain the trends
(the trends were determined from fluxes, not magnitudes).
The result in figure \ref{fig:stchaave} did not show
a strong tendency that the object was brighter during
IW And-type states, although a very faint state
(BJD 2458600--2458700) corresponded to a long-lasting dwarf
nova-type state.

\section*{Acknowledgements}

This work was supported by JSPS KAKENHI Grant Number 21K03616.
This paper includes data collected with the TESS mission, 
obtained from the MAST data archive at the Space Telescope 
Science Institute (STScI). Funding for the TESS mission is
provided by the NASA Explorer Program. STScI is operated
by the Association of Universities for Research in Astronomy,
Inc., under NASA contract NAS 5-26555.
We are grateful to the ASAS-SN team for
for making the database available to the public.

\newcommand{\noop}[1]{}\newcommand{\hyphalt}{-}



\begin{thebibliography}{}

\bibitem[{Cleveland}(1979)]{LOWESS}
  {Cleveland}, W.~S. (1979) {Robust locally weighted regression and smoothing
  scatterplots}. {\em J. Amer. Statist. Assoc.\,} {\bf 74}, 829

\bibitem[{Fernie}(1989)]{fer89error}
  {Fernie}, J.~D. (1989) Uncertainties in period determinations. {\em PASP\,}
  {\bf 101}, 225

\bibitem[{Hellier}(2000)]{hel00swsexreview}
  {Hellier}, C. (2000) The {SW Sextantis} stars. {\em New\ Astron.\ Rev.\,}
  {\bf 44}, 131

\bibitem[{Kato}(2019)]{kat19iwandtype}
  {Kato}, T. (2019) Three {Z Cam}-type dwarf novae exhibiting {IW And}-type
  phenomenon. {\em PASJ\,} {\bf 71}, 20

\bibitem[{Kato}(2021a)]{kat21v363lyr}
  {Kato}, T. (2021a) {\noop{Kato2021vsolj85}}{Periodic} modulations during a
  long outburst in {V363 Lyr}. {\em VSOLJ\ Variable\ Star\ Bull.\,} {\bf 85},
  (arXiv:2111.07237)

\bibitem[{Kato}(2021b)]{kat21ixvel}
  {Kato}, T. (2021b) {\noop{Kato2021vsolj87}}{Study} of the low-amplitude {Z
  Cam} star {IX Vel}. {\em VSOLJ\ Variable\ Star\ Bull.\,} {\bf 87},
  (arXiv:2111.15145)

\bibitem[{Kato} and {Hambsch}(2021)]{kat21stcha}
  {Kato}, T., \& {Hambsch}, F.-J. (2021) On the nature of embedded precursors
  in long outbursts of {SS Cyg} stars as inferred from observations of the {IW
  And} star {ST Cha}. {\em VSOLJ\ Variable\ Star\ Bull.\,} {\bf 83},
  (arXiv:2110.10321)

\bibitem[{Kato} et~al.(2010)]{Pdot2}
  {Kato}, T. {et~al.} (2010) {Survey of Period Variations of Superhumps in {SU
  UMa}-Type Dwarf Novae. {II}. The Second Year (2009-2010)}. {\em PASJ\,} {\bf
  62}, 1525

\bibitem[{Kimura} et~al.(2020)]{kim20iwandmodel}
  {Kimura}, M., {Osaki}, Y., {Kato}, T., \& {Mineshige}, S. (2020)
  Thermal-viscous instability in tilted accretion disks: toward understanding
  {IW And}-type dwarf novae. {\em PASJ\,} {\bf 72}, 22

\bibitem[{Kochanek} et~al.(2017)]{koc17ASASSNLC}
  {Kochanek}, C.~S. {et~al.} (2017) {The All-Sky Automated Survey for
  Supernovae} ({ASAS-SN}) light curve server v1.0. {\em PASP\,} {\bf 129},
  104502

\bibitem[{Mauder} and {Sosna}(1975)]{mau75stcha}
  {Mauder}, H., \& {Sosna}, F.~M. (1975) Light changes of {T Tauri} stars in
  the {Chamaeleon} {Association}. {\em IBVS\,} {\bf 1049}

\bibitem[{Shappee} et~al.(2014)]{ASASSN}
  {Shappee}, B.~J. {et~al.} (2014) The man behind the curtain: {X}-rays drive
  the {UV} through {NIR} variability in the 2013 {AGN} outburst in {NGC 2617}.
  {\em ApJ\,} {\bf 788}, 48

\bibitem[{Simonsen}(2011)]{sim11zcamcamp1}
  {Simonsen}, M. (2011) The {Z CamPaign}: Year 1. {\em J.\ American\ Assoc.\
  Variable\ Star\ Obs.\,} {\bf 39}, 66

\bibitem[{Simonsen} et~al.(2014)]{sim14stchabpcra}
  {Simonsen}, M., {Bohlsen}, T., {Hambsch}, F.-J., \& {Stubbings}, R. (2014)
  {ST Chamaeleontis} and {BP Coronae Australis}: Two southern dwarf novae
  confirmed as {Z Cam} stars. {\em J.\ American\ Assoc.\ Variable\ Star\
  Obs.\,} {\bf 42}, 199

\bibitem[{Steiner} et~al.(1988)]{ste88stcha}
  {Steiner}, J.~R., {Cieslinski}, D., \& {Jablonski}, F.~J. (1988) in ASP\
  Conf.\ Ser.\ 1, {Progress and Opportunities in Southern Hemisphere Optical
  Astronomy. The CTIO 25th Anniversary Symposium}, ed. V.~M. {Blanco}, \& M.~M.
  {Phillips} (San Francisco: ASP) p.~67

\bibitem[Stellingwerf(1978)]{PDM}
  Stellingwerf, R.~F. (1978) Period determination using phase dispersion
  minimization. {\em ApJ\,} {\bf 224}, 953

\bibitem[Szkody and Mattei(1984)]{szk84AAVSO}
  Szkody, P., \& Mattei, J.~A. (1984) Analysis of the {AAVSO} light curves of
  21 dwarf novae. {\em PASP\,} {\bf 96}, 988

\bibitem[Warner(1995)]{war95book}
  Warner, B. (1995) Cataclysmic Variable Stars (Cambridge: Cambridge University
  Press)

\bibitem[{Zasche} and {Paschke}(2012)]{zas12hshya}
  {Zasche}, P., \& {Paschke}, A. (2012) {HS Hydrae} about to turn off its
  eclipses. {\em A\&A\,} {\bf 542}, L23

\end{thebibliography}
\end{document}